\documentstyle[aps,multicol,epsf]{revtex}

\newcommand{\be}{\begin{equation}}
\newcommand{\ee}{\end{equation}}
\newcommand{\bea}{\begin{eqnarray}}
\newcommand{\eea}{\end{eqnarray}}
\newcommand{\ba}{\begin{array}}
\newcommand{\ea}{\end{array}}

\begin{document}
\title{
  \begin{flushright} \begin{small}
    hep-th/9912192
  \end{small} \end{flushright}
\vspace{1.cm}
Bianchi Type I Cosmology in $N=2, D=5$ Supergravity}
\author{Chiang-Mei Chen\footnote{E-mail: cmchen@joule.phy.ncu.edu.tw}}
\address{Department of Physics, National Central University,
         Chungli 320, Taiwan}
\author{T. Harko\footnote{E-mail: tcharko@hkusua.hku.hk}
    and M. K. Mak\footnote{E-mail: mkmak@vtc.edu.hk}}
\address{Department of Physics, University of Hong Kong,
         Pokfulam, Hong Kong}
\maketitle

\begin{abstract}
The dynamics and evolution of Bianchi type I space-times is considered
in the framework of the four-dimensional truncation of a reduced theory
obtained from the $N=2,D=5$ supergravity.
The general solution of the gravitational field equations can be
represented in an exact parametric form.
All solutions have a singular behavior at the initial/final moment,
except when the space-time geometry reduces to the isotropic flat case.
Generically the obtained cosmological models describe an anisotropic,
expanding or collapsing, singular Universe with a non-inflationary
evolution for all times.
\end{abstract}
\vspace{.5cm}
\hspace{1cm}{PACS number(s): 04.20.Jb, 04.65.+e, 98.80.-k}

\begin{multicols}{2}
\narrowtext
\section{Introduction}
The similarity between the $D=5$ simple supergravity (SUGRA) and $D=11$
SUGRA has been recognized for a long time \cite{Cr81,ChNi80}.
The $D=11$ SUGRA is supposed to play a fundamental role as the
low energy limit of the M-theory \cite{Wi95} --- an expected unified
speculation for the well-known five consistent superstring theories.
The field contents of the $D=11$ SUGRA theory consist of the metric,
a single Majorana spin-$\frac32$ fermion along with a (singlet)
three-form gauge potential, with neither ``$N>1$'' extensions
nor matter coupling permitted \cite{CrJuSc78}.
The simple $D=5$ SUGRA, besides the metric $\hat g_{AB}$, contains a
spin-$\frac32$ field $\hat \Psi^a_A$ ($a=1,2$ is an internal index)
and $U(1)$ gauge field (one-form $\hat B_A$) which replaces the
three-form gauge field in the $D=11$ SUGRA.
The ``primeval'' likeness comes directly from the fact that
the Lagrangians of both SUGRAs are exactly of the same form,
except for the numbering of the gauge field indices.
In addition, their dimensional reduction to $D=4$ can be carried out in
a similar way \cite{CrJu79}. Furthermore,
the $D=5$ simple SUGRA can be realized as a Calabi-Yau compactification
of the $D=11$ SUGRA together with the truncation of the scalar multiplets,
which is always necessary since there arises at least one scalar
multiplet for any Calabi-Yau compactification
\cite{CaCeAuFe95,FeKhMi96}.
Further resemblances between the two SUGRAs are related to the duality
groups upon dimensional reduction and the world sheet structure of the
solitonic string of the $D=5$ SUGRA \cite{MiOh98}.

Thus the four-dimensional reduced effective action of the $N=2, D=5$
SUGRA contains an additional Maxwell-like $U(1)$ field and a scalar
field regarded as external fields in five dimensions which are contributed
by $\hat B$, besides the ones coming from the metric
$\hat g_{AB}$ as in the traditional scheme for the Kaluza-Klein theory
\cite{AuFrMaRe82,AuMaReFr81,BaFaKe90a,BaFaKe90b}.
Cosmological solutions to this model have been previously considered
by Balbinot, Fabris and Kerner \cite{BaFaKe90a,BaFaKe90b}.
For the case of spatial homogeneity and isotropy the general
solution is non-singular in the scale factor, but unstable due to the
collapse to zero of the size of the fifth dimension
\cite{BaFaKe90a,BaFaKe90b}.
Biaxial (with two equal scale factors) anisotropic solutions with a
cylindrical homogeneous five-dimensional metric lead to singular
solutions with positive gravitational coupling \cite{BaFaKe90b}.
Recently, an explicit example of a manifestly $U$-duality covariant
M-theory cosmology in five dimensions resulting from compactification
on a Calabi-Yau three-fold has been obtained in \cite{LuOvWa98}.
Exact static solutions in $N=2,D=5$ SUGRA have been found by Pimentel
\cite{Pi95}, in a metric with cylindrical symmetry, with a particular
case corresponding to the exterior of a cosmic string.

The purpose of the present paper is to construct the general solution
to the gravitational field equations of the $N=2, D=5$ SUGRA as
formulated in \cite{BaFaKe90a,BaFaKe90b} for an anisotropic triaxial
(all directions have unequal scale factors) Bianchi type I space-time.
In this case the general solution of the field equations can be
expressed in an exact parametric form.
For all cosmological solutions, the singularity at the starting/ending
time of the evolution can not be avoided except in the isotropic limit
considered in \cite{BaFaKe90a,BaFaKe90b}.
Nevertheless, in the models analyzed in this paper, the anisotropic
Universe has non-inflationary evolution for all times and for
all values of parameters.

The present paper is organized as follows.
The field equations of our model are written down in Section II.
In Section III the general solution of the field equations is obtained.
We discuss our results and conclusions in Section IV.

\section{Field Equations, Geometry and Consequences}
The bosonic sector of $N=2, D=5$ SUGRA contains the five-dimensional
metric $\hat g_{AB}$ and $U(1)$ gauge field $\hat B_A$ described by
a Lagrangian which possesses a non-vanishing Chern-Simons term \cite{Cr81}
\bea \label{L5}
\hat {\cal L} &=& \sqrt{-\hat g} \left\{ \hat R
   - \frac14 \hat F_{AB} \hat F^{AB} \right\} \nonumber \\
  &-& \frac1{12\sqrt{3}} \epsilon^{ABCDE}\hat F_{AB}\hat F_{CD}\hat B_E,
\eea
where $\hat F_{AB} = 2\partial_{[A} \hat B_{B]}$.
In this paper we use the following conventions and notations.
The variables with hats are five-dimensional objects all other variables
are four-dimensional. Upper case indices $A,B,...$ are
used for five-dimensional space-time, greek indices
$\mu,\nu,...$ and low case indices $i,j,...$ are for
four-dimensional space-time and three-dimensional space
respectively. The signature is $(-,+,+,+,+)$.

Assuming that the five dimensional space-time has locally the structure
of $M^4 \times S^1$ with a four-dimensional space-time $M^4$
whose spatial sections are homogeneous and asymptotic flat, then
the five-dimensional metric can be decomposed along the standard
Kaluza-Klein pattern
\be
d \hat s^2 = \phi^2 (dx_4 + A_\mu dx^\mu)^2 + g_{\mu\nu} dx^\mu dx^\nu,
\ee
where the scale factor $\phi$ and Kaluza-Klein vector $A_\mu$ are
functions depending on $x^\mu$ only.

Looking for a ``ground state'' configuration we set, following
\cite{BaFaKe90a,BaFaKe90b}, the Kaluza-Klein vector, $A_\mu$, equal
to zero and take the one-form potential $\hat B_A$ to be $\hat B_\mu=0$
and $\hat B_4= \sqrt{3} \psi(x^\mu)$.
Under this ans\"atz, the five-dimensional gravitational field equations
for (\ref{L5}) reduce to a set of four-dimensional equations
\bea
R_{\mu\nu} &-& \phi^{-1} D_\mu D_\nu \phi \nonumber\\
  &-& \frac12 \phi^{-2} \left[ 3 \partial_\mu \psi \partial_\nu \psi
   - g_{\mu\nu} (\partial\psi)^2 \right] = 0, \label{FEA1} \\
D^2 \phi &+& \phi^{-1} (\partial\psi)^2 = 0, \label{FEA2} \\
D^2 \psi &-& \phi^{-1} \partial_\mu \phi \partial^\mu \psi = 0,
    \label{FEA3}
\eea
where $D$ denotes the four-dimensional covariant derivative with
respect to the metric $g_{\mu\nu}$.
Equivalently, the field equations can be re-derived,
in the string frame, from the four-dimensional Lagrangian
\cite{BaFaKe90a,BaFaKe90b}
\be \label{L4}
{\cal L} = \sqrt{-g} \phi \left\{ R
   - \frac32 \phi^{-2} (\partial\psi)^2 \right\},
\ee
via variation with respect to the fields $g_{\mu\nu}, \phi$ and $\psi$.
In the Lagrangian (\ref{L4}), the scale factor $\phi$ is an analogue
of the Brans-Dicke field whereas the origin of $\psi$ is purely
supersymmetric.

The line element of an anisotropic homogeneous flat Bianchi type I
space-time is given by
\be
ds^2 = - dt^2 + a_1^2(t) dx^2 + a_2^2(t) dy^2 + a_3^2(t) dz^2.
\ee
Defining the ``volume scale factor'', $V := \prod_i a_i$,
``directional Hubble factors'', $H_i := \dot a_i/a_i$,
and ``average Hubble factor'', $H := \frac13 \sum_i H_i$, one can
promptly find the relation $3H = \dot V/V$,
where dot means the derivative with respect to time $t$.
In terms of those variables, the field equations
(\ref{FEA1}) and the equations of motion for $\phi$ and $\psi$
(\ref{FEA2},\ref{FEA3}) coupling with the anisotropic Bianchi type I
geometry take the concise forms
\bea
3 \dot H + \sum_i H_i^2 + \phi^{-1} \ddot \phi
   + \phi^{-2} \dot \psi^2 &=& 0, \label{FEB1} \\
V^{-1} \frac{d}{dt} (VH_i) + H_i \phi^{-1} \dot\phi
   - \frac12 \phi^{-2} \dot \psi^2 &=& 0, \; i=1,2,3,
     \label{FEB2} \\
V^{-1} \frac{d}{dt} (V\dot\phi) + \phi^{-1} \dot \psi^2
  &=& 0, \label{FEB3} \\
V^{-1} \frac{d}{dt} (V\dot\psi) - \phi^{-1} \dot\phi \dot\psi\
  &=& 0. \label{FEB4}
\eea

The physical quantities of interest in cosmology are the {\em expansion
scalar} $\theta$, the {\em mean anisotropy parameter} $A$,
the {\em shear scalar} $\sigma^2$ and the {\em deceleration parameter}
$q$ defined as \cite{Gr85}
\bea
\theta &:=& 3H, \qquad
A := \frac13 \sum_i \left( \frac{H-H_i}{H} \right)^2, \nonumber \\
\sigma^2 &:=&
     \frac12 \left( \sum_i H_i^2 - 3 H^2 \right), \quad
q := \frac{d}{dt} H^{-1} - 1. \label{Def}
\eea

The sign of the deceleration parameter indicates whether the cosmological
model inflates. A positive sign corresponds to standard decelerating
models whereas a negative sign indicates inflationary behavior.

\section{General Solution of the Field Equations}
Equation (\ref{FEB4}) can immediately be integrated to give
\be
V \dot \psi = \omega \phi, \label{VPP}
\ee
with $\omega$ --- a constant of integration.
From equations (\ref{FEB3}) and (\ref{FEB4}) one can find that the
expressions of the fields $\phi(t)$ and $\psi(t)$ have the following form
\bea
\phi(t) &=& \phi_0 \cos \left( \omega \int \frac{dt}{V}
   + \omega_0 \right), \label{phi} \\
\psi(t) &=& \psi_0 + \phi_0 \sin \left( \omega \int \frac{dt}{V}
   + \omega_0 \right), \label{psi}
\eea
where $\phi_0, \psi_0$ and $\omega_0$ are constants of integration.

By summing equations (\ref{FEB2}) one gets
\be
V^{-1} \frac{d}{dt} (VH) + H \phi^{-1} \dot\phi
   - \frac12 \phi^{-2} \dot \psi^2 = 0, \label{FEB2a}
\ee
which can be transformed, by using equations (\ref{VPP}) and (\ref{phi}),
into the following differential-integral equation
describing the dynamics and evolution
of a triaxial Bianchi type I space-time in $N=2,D=5$ SUGRA:
\be \label{EqV}
\ddot V = \omega \frac{\dot V}{V} \tan \left( \omega \int \frac{dt}{V}
   + \omega_0 \right) + \frac32 \frac{\omega^2}{V}.
\ee

Furthermore, by subtracting equation (\ref{FEB2a}) from equations
(\ref{FEB2}), one can solve for the $H_i$ as
\be
H_i = H + \frac{K_i}{\phi V}, \qquad i=1,2,3, \label{Hi}
\ee
where $K_i$ are constants of integration satisfying the following
consistency condition
\be
\sum_i K_i = 0. \label{Ki}
\ee
Therefore the physical quantities of interest (\ref{Def}) reduce to
\be
A = \frac{K^2}{3\phi^2 V^2 H^2}, \qquad
\sigma^2 = \frac32 A H^2,
\ee
where $K^2 = \sum_i K_i^2$.

By introducing a new variable $\eta$ related to the physical time $t$
by means of the transformation $d\eta := dt/V$ and by denoting
$u:=\dot V = dV/(V d\eta)$, equation (\ref{EqV}) reduces to a first
order linear differential equation for the unknown function $u$
\be
\frac{du}{d\eta} = \omega \tan( \omega\eta + \omega_0) u
   + \frac32 \omega^2,
\ee
whose general solution is given by
\be
u = C \cos^{-1}(\omega\eta+\omega_0)
  + \frac32 \omega \tan(\omega\eta+\omega_0), \label{dotV}
\ee
where $C$ is an arbitrary constant of integration.

Defining a new parameter $\zeta(\eta):=\omega\eta+\omega_0$,
we can represent the general solution of
the field equations for a Bianchi type I space-time in
the $N=2,D=5$ SUGRA in the following exact parametric form:
\bea
t &=& t_0 + \frac{V_0}{\omega} \int
   \frac{(1+\sin\zeta)^\beta}{(1-\sin\zeta)^\gamma} d\zeta,
   \label{T} \\
V &=& V_0 \frac{(1+\sin\zeta)^\beta}{(1-\sin\zeta)^\gamma},
   \label{V} \\
H &=& \frac{\omega}{2V_0} (\alpha + \sin\zeta)
   \frac{(1-\sin\zeta)^{\gamma-\frac12}}{(1+\sin\zeta)^{\beta+\frac12}}, \\
a_{i} &=& a_{i0}
   \frac{(1+\sin\zeta)^{\frac{\beta}3+\frac{K_i}{2\omega\phi_0}}}
        {(1-\sin\zeta)^{\frac{\gamma}3+\frac{K_i}{2\omega\phi_0}}},
   \quad i=1,2,3, \label{ai}
\eea
where we have denoted $\alpha = \frac{2C}{3\omega}$,
$\beta=\frac34(\alpha-1)$, $\gamma=\frac34(\alpha+1)$ and the $a_{i0}$
are arbitrary constants of integration while $V_0=\Pi_i a_{i0}$.
The observationally important physical quantities are given by
\bea
A &=& \frac{4K^2}{3\phi_0^2 \omega^2}
      \left( \alpha + \sin\zeta \right)^{-2}, \label{AA} \\
\sigma^2 &=& \frac{K^2}{2\phi_0^2 V_0^2}
   \frac{(1-\sin\zeta)^{2\gamma-1}}{(1+\sin\zeta)^{2\beta+1}}, \\
q &=& 2 \left\{ 1 - \frac{ 1 + \alpha \sin\zeta}
                         {(\alpha + \sin\zeta)^2} \right\}. \label{qq}
\eea
Finally, the field equation (\ref{FEB1}) gives a
consistency condition relating the constants $K^2, \omega, \alpha$ and
$\phi_0$:
\be
K^2 = \frac32 \phi_0^2 \omega^2 \left( \alpha^2 - 1 \right), \label{K2}
\ee
leading to
\be
\alpha \ge 1 \quad \hbox{or} \quad \alpha \le -1.
\ee

It is worth noting that these two classes of solutions corresponding to
positive or negative values of $\alpha$ and $\omega$ are not independent.
Indeed, they can be related via a ``duality'' transformation by
changing the signs of $\omega, \alpha$ and $\zeta$ so that
$t(\omega,\alpha,\zeta)=t(-\omega,-\alpha,-\zeta)$,
$a_i(\omega,\alpha,\zeta)=a_i(-\omega,-\alpha,-\zeta), i=1,2,3$,
and $V(\alpha,\zeta)=V(-\alpha,-\zeta)$ etc.
This duality relation can be obtained by a simple inspection of equations
(\ref{T})-(\ref{ai}) and, therefore, all physical quantities are
invariant with respect to this transformation.
Moreover, the physical properties of the cosmological
models presented here are strongly dependent on the signs of the
parameters $\alpha$ and $\omega$.
Nevertheless, due to the duality transformation, hereafter we will consider,
without loss of generality, the cases with positive $\alpha$ only.

For some particular values of $\alpha$, the general solutions
can be expressed in an exact non-parametric form, for instance,
an exact class solutions can be obtained for $\alpha = \pm \frac53$.
By introducing a new time variable $\tau:=\frac{3\sqrt{2}\omega}{V_0}t$,
and choosing $t_0=\mp\frac{V_0}{3\sqrt{2}\omega}$,
the exact solution in $N=2,D=5$ SUGRA
for the Bianchi type I space-time is given by
\bea
\tau &=& \pm \left[ \cos^{-3} \left( \frac{\zeta}2 \pm\frac{\pi}4 \right)
    -1 \right], \\
V &=& \frac{V_0}{2\sqrt{2}} (1\pm\tau) \left[ (1\pm\tau)^\frac23 - 1
    \right]^\frac12, \\
H &=& \frac{\sqrt{2}\omega}{V_0} \frac{\frac43 (1\pm\tau)^\frac23 -1}
    {(1\pm\tau)\left[ (1\pm\tau)^\frac23 - 1\right]}, \\
a_i &=& \frac{a_{i0}}{\sqrt{2}} (1\pm\tau)^\frac13 \left[
    (1\pm\tau)^\frac23 - 1 \right]^{\frac16\pm\frac{K_i}{2\omega\phi_0}}, \\
A &=& \frac89 (1\pm\tau)^\frac43 \left[ \frac43 (1\pm\tau)^\frac23 - 1
    \right]^{-2}, \\
\sigma^2 &=& \frac{8\omega^2}{3V_0^2} (1\pm\tau)^{-\frac23}
    \left[ (1\pm\tau)^\frac23 - 1 \right]^{-2}, \\
q &=& 2 \left\{ 1-\frac{(1\pm\tau)^\frac23\left[4(1\pm\tau)^\frac23-5\right]}
    {6\left[\frac43(1\pm\tau)^\frac23-1\right]^2} \right\}.
\eea

The isotropic limit can be achieved by taking $\alpha=\pm 1$
and, consequently, $K_i=0,i=1,2,3$.
It is worth noting that our solutions reduce to two different
types of homogeneous space-times when $\alpha=\pm 1$.

For $\alpha=1$, we obtain (by denoting $a_1=a_2=a_3=a$)
\bea
t &=& t_0+\frac{V_0}{2\sqrt{2}\omega}\left[ \frac{\sin\theta}{\cos^2\theta}
    + \ln(\tan\theta+\sec\theta) \right], \label{IT1} \\
a &=& \frac{a_0}{2\sqrt{2}} \cos^{-3}\theta, \label{IA1}
\eea
where $\theta:=\frac{\zeta}2 + \frac{\pi}4$.
Eqs.(\ref{IT1}) and (\ref{IA1}), describing a homogeneous flat isotropic
space-time interacting with two scalar fields (Kaluza-Klein and
supersymmetric), have been previously obtained by Balbinot, Fabris and
Kerner \cite{BaFaKe90a} (for an extra choice of the parameter
$\omega_0=\pi/2$), who extensively studied their physical properties.
This isotropic solution also provides a positive gravitational coupling
at the present time.

In the isotropic limit corresponding to $\alpha=-1$, one can obtain
another class of isotropic homogeneous flat space-times represented in
the following parametric form by
\bea
t &=& t_0-\frac{V_0}{2\sqrt{2}\omega}\left[ \frac{\cos\theta}{\sin^2\theta}
    - \ln(\csc\theta-\cot\theta) \right], \label{IT2} \\
a &=& \frac{a_0}{2\sqrt{2}} \sin^{-3}\theta.  \label{IA2}
\eea
This type of flat space-time has not been previously considered.

\section{Discussions and Final Remarks}
In order to study the physical properties of the Bianchi type I Universe
described by the Eqs. (\ref{V})-(\ref{ai}) we need to fix first the range
of variation of the parameter $\zeta$. There are no a priori limitations
in choosing the admissible range of values, thus both positive and negative
values are permitted since the variable $\eta=\int dt/V$ can also
be negative.
But from a physical point of view it is natural to impose the condition
such that the gravitational coupling $\phi$ is always positive during
the evolution of the Bianchi type I space-time in $N=2,D=5$ SUGRA.
Consequently, we shall consider $\zeta\in (-\pi/2, \pi/2)$.
With this choice, the Universe for $\alpha<-1$ starts its evolution
in the infinite past ($t\to -\infty$) and ends at a finite moment $t=t_0$.
For $\alpha>1$ the Universe starts at $t=t_0$ and ends in an infinite
future with $t \to \infty$. (All discussion here and hereafter are with
respect to positive $\omega$.)

As can be easily seen from equation (\ref{dotV}), if $\alpha<-1$
we have $\dot V < 0$ for all $t$.
For these values of the parameters the
Bianchi type I anisotropic Universe collapses from an initial state
characterized by infinite values of the volume scale factor and of the
scale factors $a_i,i=1,2,3,$ to a singular state
with all factors vanishing.
But if $\alpha>1$, the Universe expands and $\dot V > 0$ for all $t$.
The expanding Bianchi type I Universe starts its evolution at the
initial moment, $t_0$, from a singular state with zero values of the
scale factors, $a_i(t_0)=0,i=1,2,3$.

Another possible way to investigate the singularity behavior at the
initial moment is to consider the sign of the quantity
$R_{\mu\nu} u^\mu u^\nu$, where $u^\mu$ is the vector tangent to the
geodesics; $u^\mu=(-1,0,0,0)$ for the present model.
From the gravitational field equations we easily obtain
\be
R_{\mu\nu} u^\mu u^\nu = 3 H^2 (q-A). \label{Ruu}
\ee
By using equations (\ref{AA}), (\ref{qq}) and (\ref{K2}) we can
express (\ref{Ruu}) as
\be
R_{\mu\nu} u^\mu u^\nu
    = 6H^2 \frac{\sin\zeta}{\alpha+\sin\zeta}.
\ee
For $\zeta \to -\pi/2$ the sign of $R_{\mu\nu}u^\mu u^\nu$ is determined
by the sign of $1-\alpha$.
Therefore we obtain
\be
R_{\mu\nu} u^\mu u^\nu < 0 \quad \hbox{for} \quad \alpha > 1.
\ee
Hence, the energy condition of Hawking-Penrose singularity theorems
\cite{HaEl73} is not satisfied for the solutions corresponding to Bianchi
type I Universes in the four-dimensional reduced two scalar fields theory of
$N=2,D=5$ SUGRA with $\alpha>1$.
Nevertheless, for those solutions an initial singular state is
{\em unavoidable} at the initial moment $t_0$.

Since for $\alpha>1$ the Bianchi type I Universe starts its evolution at
the initial moment $t=t_0$ ($\zeta \to -\pi/2$) from a singular state,
therefore, the presence of a variable gravitational coupling, $\phi$,
and of a supersymmetric field, $\psi$, in an anisotropic geometry {\em can
not} remove the initial singularity that mars the big-bang cosmology.
At the initial moment the degree of the anisotropy of the space-time
is maximal, with the initial value of the anisotropy
parameter $A(t_0)=2(\alpha+1)/(\alpha-1)$.
For $t>t_0$ the Universe expands and the anisotropy parameter decreases.

\begin{figure}
\epsfxsize=10cm
\epsffile{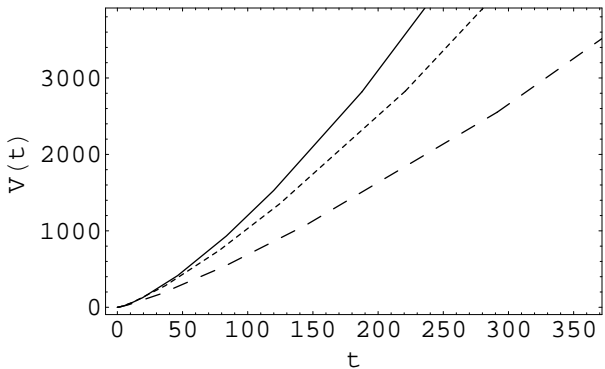}
\caption{Behavior of the volume scale factor of the Bianchi type I
         space-time for different values of $\alpha>1$
         ($V_0=1$ and $\omega=1$): $\alpha=3/2$ (solid curve),
         $\alpha=2$ (dotted curve) and $\alpha=5/2$ (dashed curve).}
\label{FIG1}
\end{figure}

\begin{figure}
\epsfxsize=10cm
\epsffile{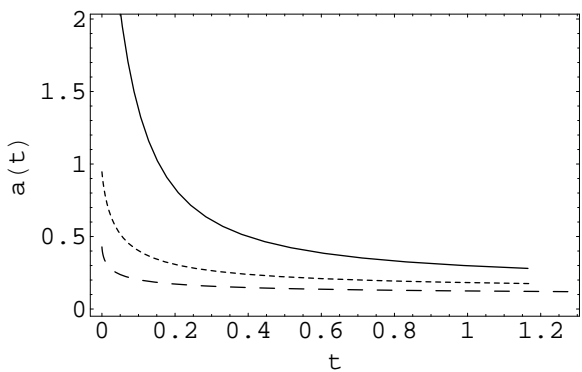}
\caption{Time dependence of the parameter
         $a=\frac{3\phi_0^2\omega^2}{4K^2} A$ for different values of
         $\alpha$: $\alpha=3/2$ (solid curve),
         $\alpha=2$ (dotted curve) and $\alpha=5/2$ (dashed curve).}
\label{FIG2}
\end{figure}

\begin{figure}
\epsfxsize=10cm
\epsffile{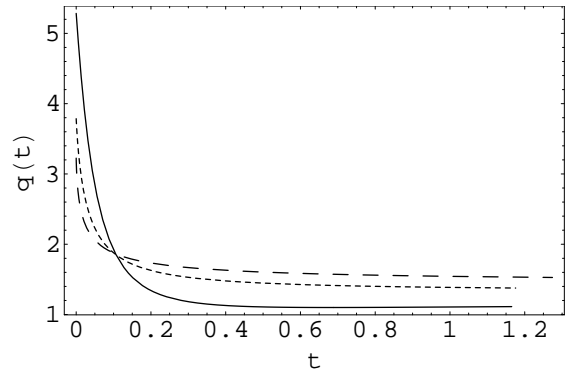}
\caption{Evolution of the deceleration parameter $q$ of the Bianchi
         type I space-time for different values of $\alpha$:
         $\alpha=3/2$ (solid curve),
         $\alpha=2$ (dotted curve) and $\alpha=5/2$ (dashed curve).}
\label{FIG3}
\end{figure}

The behavior of the volume scale factor, of the anisotropy parameter
and of the deceleration parameter is presented for different values of
$\alpha$ in Figs. \ref{FIG1}-\ref{FIG3}.
The evolution of the Universe is non-inflationary, with $q>0$
for all $t>t_0$. Non-inflationary behavior is a generic feature of most
of the supersymmetric models.
This is due to the general fact that the effective potential for the
inflation field $\sigma$ in SUGRA typically is too curved,
growing as $\exp (c \sigma^2/m)$, with $c$ a parameter
(typically $O(1)$) and $m$ the stringy Planck mass \cite{LiRi97}.
The typical values of $c$ make inflation impossible because the
inflation mass becomes of the order of the Hubble constant.
See also \cite{LiRi97} for a simple realization of hybrid inflation
in SUGRA.
In the present model the presence of the supersymmetric field $\psi$
prevents the Bianchi type I Universe from inflating.

\begin{figure}
\epsfxsize=10cm
\epsffile{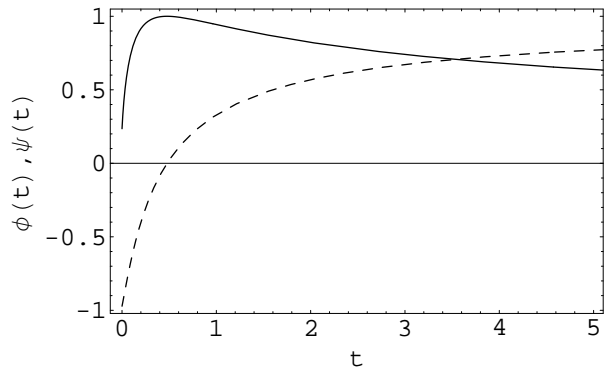}
\caption{Time evolution of the gravitational coupling $\phi$ (solid curve)
         and of the supersymmetric field $\psi$ (dashed curve)
         for $\alpha=5/2$ ($\phi_0=1$, $\psi_0=0$).}
\label{FIG4}
\end{figure}

In the far future, for $\zeta \to \pi/2$ and $t\to \infty$ ($\alpha>1$),
we have $V \to \infty, a_i \to\infty, i=1,2,3$.
In this limit the anisotropy parameter becomes a non-zero constant and
the Universe ends in a still anisotropic phase, but with
a decrease in the value of the anisotropy parameter $A$,
as compared with the initial one.
Therefore during its evolution the Bianchi type I Universe cannot
experience a transition from the anisotropic phase to the isotropic
flat geometry.
The time evolution of the gravitational coupling $\phi$ and of the
supersymmetric field $\psi$ is represented in Fig. \ref{FIG4}.
The $\phi$-field is positive for all values of time.

In the present paper we have investigated the evolution and dynamics
of a Bianchi type I space-time in a SUGRA toy-model, obtained
by dimensional reduction of the $N=2,D=5$ SUGRA.
The inclusion of the supersymmetric term gives some particular
features to this cosmological model, by preventing the Universe from
inflating and attaining completely isotropy.
But globally there is a decrease in the degree of anisotropy of the
geometry. Hence this model can be used to describe only a specific,
well-determined period of the evolution of our Universe.

\section*{Acknowledgments}
One of the authors (CMC) would like to thank prof. J.M. Nester for
profitable discussions.
The work of CMC was supported in part by the National Science Council
(Taiwan) under grant NSC 89-2112-M-008-016.

We are also grateful to prof. Pimentel for calling our attention to the
results \cite{Pi92,PiSo93,PiSo95} about several types of Bianchi
cosmologies in the framework of $N=2,D=5$ supergravity.

\begin{appendix}
\section{Some Exact Forms for Physical Time}
For a large class of values of the parameter $\alpha$ the general
solution of the gravitational field equations can be expressed in a
closed explicit form.
The variation of the physical time $t$ is determined by the integral
equation (\ref{T}). After a trick manipulation, one can rewrite this
equation in the following form
\be
t = t_0 + \frac{V_0}{\sqrt{2}\omega} \int \sin^{\gamma'-3}\theta
    \cos^{-\gamma'}\theta d \theta,
\ee
where $\theta:=\frac{\zeta}2+\frac{\pi}4$ and
$\gamma':=2\gamma=\frac32(\alpha+1)$.
In general, for arbitrary $\gamma'$, this integral can not be closed.
Fortunately, for integer values of $\gamma'$, the physical time $t$ can
be expressed in an explicit form as a function of $\zeta$. Some of these
exact forms of the time function are listed in the following.
(The outcomes for $\alpha\pm1$ are given in (\ref{IT1}) and (\ref{IT2}) ).

\noindent{\bf For $\alpha < -1$ :}

\noindent{(i) $\alpha=-\frac53, \; (\gamma'=-1),$}
$$ t=t_0-\frac{V_0}{3\sqrt{2}\omega} \frac1{\sin^3\theta}, $$

\noindent{(ii) $\alpha=-\frac73, \; (\gamma'=-2),$}
$$
t=t_0-\frac{V_0}{8\sqrt{2}\omega}\Biggl[
     \frac{\cos\theta(\cos^2\theta+1)}{\sin^4\theta}
   + \ln(\csc\theta-\cot\theta) \Biggr],
$$

\noindent{(iii) $\alpha=-3, \; (\gamma'=-3),$}
$$
t=t_0-\frac{V_0}{15\sqrt{2}\omega}\Biggl(
     \frac{5\cos^2-2}{\sin^5\theta} \Biggr).
$$

\noindent{\bf For $\alpha > 1$ :}

\noindent{(i) $\alpha=\frac53, \; (\gamma'=4),$}
$$ t=t_0+\frac{V_0}{3\sqrt{2}\omega} \frac1{\cos^3\theta}, $$

\noindent{(ii) $\alpha=\frac73, \; (\gamma'=5),$}
$$
t=t_0+\frac{V_0}{8\sqrt{2}\omega}\Biggl[
     \frac{\sin\theta(\sin^2\theta+1)}{\cos^4\theta}
   - \ln(\tan\theta+\sec\theta) \Biggr],
$$

\noindent{(iii) $\alpha=3, \; (\gamma'=6),$}
$$
t=t_0+\frac{V_0}{15\sqrt{2}\omega}\Biggl(
     \frac{5\sin^2\theta-2}{\cos^5\theta} \Biggr).
$$

\end{appendix}

\end{multicols}
\end{document}